\documentclass {aa}
\usepackage[varg]{txfonts}
\bibpunct{(}{)}{;}{a}{}{,} 
\usepackage{graphicx}

\begin{document}

\title{A Jupiter-mass planet around the K0 giant \object{HD 208897}
  \thanks{This work was supported by The Scientific and Technological Research Council of Turkey (T\"{U}B\.{I}TAK), the project number of 114F099}}

\author{M. Y{\i}lmaz\inst{1}
  \and B. Sato\inst{2}
  \and I. Bikmaev\inst{3,6}
  \and S. O. Selam\inst{1}
  \and H. Izumiura\inst{4}
  \and V. Keskin\inst{5}
  \and E. Kambe\inst{4}
  \and S. S. Melnikov\inst{3,6}
  \and A. Galeev\inst{3,6}
  \and \.{I}. \"{O}zavc{\i}\inst{1}
  \and E. N. Irtuganov\inst{3,6}
  \and R. Ya. Zhuchkov\inst{3,6,7}
       }

\offprints{M. Y{\i}lmaz, \email{mesutyilmaz@ankara.edu.tr}}

\institute{Ankara University, Department of Astronomy and Space Sciences, TR-06100, Ankara, Turkey
\and Tokyo Institute of Technology, Ookayama, Meguro-ku, Tokyo 152-8550, Japan
\and Kazan Federal University, Department of Astronomy and Satellite Geodesy, 420008, Kazan, Russia
\and Okayama Astrophysical Observatory, Honjo 3037-5, Kamogata, Asakuchi, Okayama 719-0232, Japan
\and Ege University, Department of Astronomy and Space Sciences, TR-35100, Bornova, \.Izmir, Turkey
\and Academy of Sciences of Tatarstan, Bauman Str, 20, 420111, Kazan, Russia
\and Main (Pulkovo) Astronomical Observatory, Russian Academy of Sciences, Saint-Petersburg, 196140, Russia
  }

\date{Received xx xx 2017 / Accepted xx xx 2017}

\abstract{For over 10 years, we have carried out a precise radial velocity (RV) survey to find substellar companions around evolved G,K-type stars to extend our knowledge of planet formation and evolution. We performed high precision RV measurements for the giant star \object{HD 208897} using an iodine ($I_{2}$) absorption cell.  The measurements were made at T\"{U}B\.{I}TAK National Observatory (TUG; RTT150) and Okayama Astrophysical Observatory (OAO). For the origin of the periodic variation seen in the RV data of the star, we adopted a Keplerian motion caused by an unseen companion. 
We found that the star hosts a planet with a minimum mass of $m_{2}sini=1.40M_{J}$, which is relatively low compared to those of known planets orbiting evolved intermediate-mass stars. The planet is in a nearly circular orbit with a period of $P=353$ days at about 1 AU distance from the host star. The star is metal rich and located at the early phase of ascent along the red giant branch. The photometric observations of the star at Ankara University Kreiken Observatory (AUKR) and the \textit{HIPPARCOS} photometry show no sign of variation with periods associated with the RV variation. Neither bisector velocity analysis nor analysis of the Ca II and H$\alpha$ lines shows any correlation with the RV measurements.
}

\keywords{stars: individual: \object{HD 208897} -- stars: planetary systems -- stars: fundamental parameters -- techniques: radial velocities}

\maketitle 

\section{Introduction}
The radial velocity (RV) method is a technique that is widely used to detect exoplanets through the observations of Doppler shifts of spectral lines in the spectrum of the host star of a planet. The reflex motion of a star due to a planetary companion produces a periodic variation of RV with a few $ms^{-1}$ velocity amplitude that depends on the mass and distance of the planet. However, this method requires both a high signal-to-noise (S/N) ratio and a high spectral resolution to achieve such a high RV precision. About 3500 exoplanets have been discovered with  various methods so far (The NASA Exoplanet Archive; \cite{akeson2017}), of which about 640 have been found with the RV technique. While most of the known exoplanets ($\sim$75\%) orbit around G, K-type dwarfs, a small fraction ($\sim$4\%) of these exoplanets have been found around evolved intermediate-mass ($1.3-5 M_{\odot}$) stars (e.g., \citet{niedzielski2016, ortiz2016, giguere2015, jones2015, lee2014, nowak2013, mitchell2013, sato2010, sato2013,  omiya2012, wang2012, johnson2011,  wittenmyer2011}). In spite of about 130 planets discovered around G-K giant stars, only few of these planets ($\sim20$) are close to the mass of Jupiter. Therefore, there is still a need to increase the number of planets around G-K giant to strengthen sample.

Intermediate-mass stars on the main sequence have very few absorption lines suitable for precise RV studies owing to their high atmospheric temperatures and rapid rotations. However, their evolved counterparts, giants and subgiants, are promising targets for precise Doppler-shift measurements because they show many useful spectral lines thanks to their lower surface temperatures and slower rotation rates.

Every new discovery of an exoplanet allows us not only to constrain a general picture of planet formation but also to understand how stellar evolution affects planetary systems. In particular, planets around evolved intermediate-mass stars are of great importance. Intermediate-mass stars tend to have more massive protoplanetary disks than those of sun-like stars, which provides an opportunity to investigate planet formation in different environments other than sun-like stars. At the same time, intermediate-mass stars have shorter evolutionary timescales, which causes their protoplanetary disks to also have shorter lifetimes than those of sun-like stars. In addition, the time allowed for planet formation is also much short compared to sun-like stars in intermediate-mass stars. Searches for planets around giant stars also provide a snapshot of the changes in dynamical configuration of the planetary system during evolution of the host star. In addition, surveys of evolved stars have revealed that the orbital properties of their planets seem significantly different from those of G,K dwarfs and therefore the relevant statistical outcomes \citep{adibekyan2013,bowler2010,johnson2010,takeda2008, mortier2012,pasquini2007, udry2007} are still open to debate. For example, semimajor axes of planets orbiting evolved stars are larger than 0.6 AU. Indeed, there are almost no planets orbiting closer than 0.6 AU around stars with $M>1.5M_{\odot}$ \citep{johnson2007,sato2008,wright2009}. It has been proposed that the lack of planets close to evolved intermediate-mass stars is the result of the engulfment of inner-orbit planets by the host stars when the stars ascended the red giant branch (RGB) \citep{sato2008, villaver2009}. Moreover, the planet frequency--stellar metallicity correlation in intermediate-mass stars seems tighter than that in G,K dwarfs. Furthermore, the occurrence rate of planets around intermediate-mass stars increases with increasing stellar mass \citep{reffert2015, johnson2010} and these occupy less eccentric orbits as compared to those of planets around sun-like stars \citep{johnson2008}. Therefore, the existence of planets around evolved giants leads to a test of the viability of planet formation models (e.g., \citet{boss2000, ida2004, bodenheimer1986, mordasini2008}) and evolution of planetary systems. So far, signs of various important properties have been revealed for planets in intermediate-mass stars, but they are still in need of confirmation based on a much larger number of samples.

In 2007, we started a precise Doppler survey to search for planets around evolved intermediate-mass stars using the 1.5 m Russian-Turkish Telescope (RTT150) at T\"{U}B\.{I}TAK National Observatory (TUG) within the framework of an international collaboration between Turkey, Russia, and Japan \citep{yilmaz2015}. The survey program is an extension to the ongoing Okayama Astrophysical Observatory (OAO) planet search program \citep{sato2005}. About 50 G,K-type giant stars were selected for the survey from the HIPPARCOS \citep{perryman1997} catalog according to following criteria: visual magnitude of $V \sim 6.5$, color index of $0.6 \leq B-V \leq 1.0$, declination of $\delta \geq -20^{\circ}$, and excluding stars known as photometric variables. 

In this work, we report the first planet discovery around a giant star \object{HD 208897} in our planet search program using the RTT150 and 1.88 m telescope at OAO. The paper is organized as follows: In section 2, we describe our spectroscopic observations at TUG and OAO, and we also present photometric observations at AUKR (Ankara University Kreiken Observatory) and the \textit{HIPPARCOS} \citep{vanleeuwen1997} photometry database. The stellar characteristic is presented in section 3, while orbital solutions and other possible causes of the RV variation are discussed in section 4. Finally, we present our discussions and conclusions in section 5.

\section{Observations and analysis}
Since 2007, we have been carrying out a Doppler planet search program targeting 50 G-K type giants using the RTT150 at TUG. From these observations we found that 13 targets show significant RV variations between 20 and 500 $ms^{-1}$.  Therefore we decided to follow up these targets with the 1.88 m telescope at the Okayama Astrophysical Observatory (OAO) after 2012. Moreover, we started to observe these targets photometrically at AUKR to check photometric variability or detect any transit phenomenon. One of these candidates is \object{HD 208897}.
\subsection{Observations from T\"{U}B\.{I}TAK National Observatory}
We acquired 73 spectra for \object{HD 208897} from 2009 June to 2017 January using Coude Echelle Spectrograph (CES) and $2K\times2K$ Andor CCD attached to RTT150 at TUG. We used an iodine ($I_{2}$) absorption cell in front of the entrance slit of the spectrograph to obtain precise RV measurements, which superimposes thousands of molecular absorption lines over the object spectra. Using these lines as a wavelength reference, we simultaneously derived the instrumental profile and Doppler shift relative to stellar template spectrum. The TUG CES spectra covered a wavelength region from 4000 {\AA} to 8000 {\AA} with resolving power $R\sim$55000. The typical signal-to-noise ratios ($S/N$) were obtained as 60-120 per pixel at 5500 {\AA} with an exposure time of 1800 seconds for the entire data set. The Doppler precision of RTT150 CES is about 10 $ms^{-1}$ over a time span of nine years \citep{yilmaz2015}.

\subsection{Observations from Okayama Astrophysical Observatory }
From 2014 to 2017, we used the 188 cm telescope and High Dispersion Echelle Spectrograph (HIDES) high-efficiency fiber-feeding system (hereafter HIDES-F) at OAO \citep{izumiura1999, kambe2013} and obtained a total of 34 data points for \object{HD 208897}. The HIDES-F instrument uses an image slicer as the entrance aperture of the spectrograph and its spectral resolution is fixed to $R=55000$. The spectra covered a wavelength region 3750 {\AA} to 7500 {\AA}. For precise RV measurements, we used $I_{2}$ absorption cell, which provides an ideal wavelength reference in a wavelength range of 5000 {\AA} to 5800 {\AA}. For stars with the visual magnitude of $V<6.5$ we can obtain a sufficient signal-to-noise ratio of $S/N > 200$ with an exposure time shorter than 30 min, with which the RV precision can reach down to 3 $ms^{-1}$ \citep{harakawa2015}.  

\subsection{Ankara University Kreiken Observatory Photometric observations}

Photometric observations of \object{HD 208897} were carried out with the 35 cm T35 telescope and $1K\times1K$ Apogee ALTA U47 camera at AUKR between 2014 October and 2017 January. The plate scale of camera is $0\arcsec.75$ per pixel and full field of view is $13\arcmin \times 13\arcmin$. Single color photometric data of the target were obtained using Bessel-R filter. During the observations we used the telescope defocusing technique to achieve high photometric precision by distributing the point spread function (PSF) over many pixels. This approach allowed us to minimize the effects of flat-fielding errors or seeing changes. The diameters of the defocused PSF ranged between 30 and 50 pixels.

\subsection{Data analysis}

The stellar spectra obtained at both TUG and OAO were processed following the standard echelle reduction procedures in IRAF\footnote{http://iraf.noao.edu} software packages, i.e., bias subtraction, extraction of the scattered light produced in the optical system, division by the normalized flat-field, and wavelength calibration by Thorium-Argon (ThAr) lamp reference spectra. After these reduction processes, the spectra were normalized to the continuum, order by order, by fitting a polynomial function to remove the general shape of the aperture spectra and prepare\ for the precise RV measurement procedure. The precise RVs of the target were derived from the observed star spectra taken through the $I_{2}$ cell via a custom IDL\footnote{www.harrisgeospatial.com/productsservices/idl.aspx} code for CES data and a C code for HIDES-F data, which are based on the analysis technique described by \citet{Butler1996}, \citet{Sato2002}, and \citet{Sato2012}. In this technique, we divided the echelle spectrum into hundreds of chunks with a few \AA (typically 3.5 \AA) width and applied a Doppler analysis to each chunk. The final measured RV is the weighted mean of the velocities of the individual chunks and all RVs were corrected to the solar system barycenter, which is based on the Jet Propulsion Lab (JPL) ephemeris calculations. The derived RVs are listed in Table \ref{tab1} and shown in Figure \ref{fig3} together with the estimated uncertainties.

\begin{figure}
  \includegraphics[width=0.99\columnwidth]{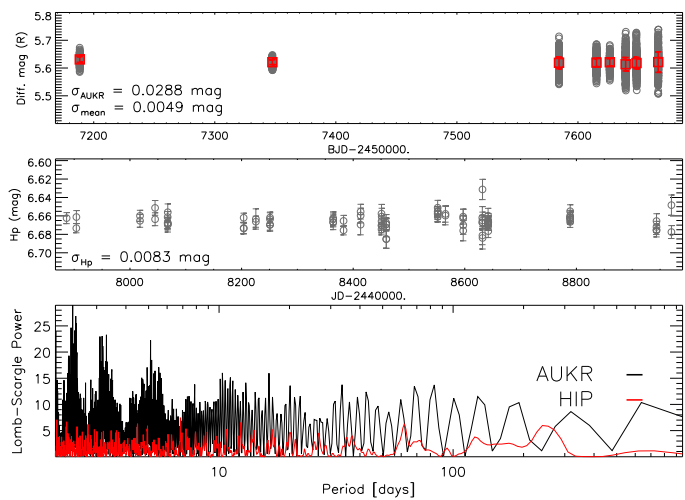}
  \caption{\textit{Top panel}: The AUKR photometric observations of \object{HD 208897}. Open red squares represent mean Bessel-R values of the individual nights. The scatter of the mean brightness is about 0.005 mag. \textit{Middle panel}: The \textit{HIPPARCOS} photometric data, indicating a photometric stability down to $\sigma \sim 0.008$ mag. \textit{Bottom panel}: Lomb-Scargle periodograms of the AUKR (black) and the \textit{HIPPARCOS} (red) photometric measurements.}
  \label{fig1}
\end{figure}

The photometric images taken AUKR were reduced via the photometric tasks of IRAF software. Then we performed standard aperture photometry with the \textit{ASTROLIB/APER} IDL routine. The magnitudes and their errors of stars within the image field were derived using this routine and after that we chose comparison stars for relative photometry. We discarded all variable or stars that are too faint from the comparison list. Finally, we obtained the relative photometry of the target by performing the weighted ensemble photometry technique for the comparison stars. The photometric data of \object{HD 208897} is comprised of 3821 measurements, spanning over eight nights and unevenly sampled. The obtained light curve is shown in the top panel of Figure \ref{fig1}. The mean AUKR photometric data exhibit a variation less than 0.005 mag, while whole data set revealed a photometric variability of $\sigma \sim 0.03$ mag. As shown in the bottom panel of Figure \ref{fig1} (black solid line), no significant periodic signal was found in the photometry of these data. In addition, we checked the \textit{HIPPARCOS} photometric variation to examine causes of the apparent RV variation other than orbital motion. The \textit{HIPPARCOS} photometry data of \object{HD 208897} obtained from December 1989 to December 1992  and consists of 84 measurements with a photometric variability of $\sigma \sim 8$ mmag (middle panel of Figure \ref{fig1}). The bottom panel of Figure \ref{fig1} shows a periodogram of the \textit{HIPPARCOS} photometry (red solid line). We could not see a significant peak at around the period of the RV variation.

\begin{table}
\caption{Radial velocities for \object{HD 208897}.} 
\scriptsize
\begin{center}
\begin{tabular}{cccc}
\hline
  BJD-2450000 [days]   & Velocity [$ms^{-1}$] & Uncertainty [$ms^{-1}$] & Remark \\
\hline
5003.52970 & -27.76 &   6.97 & TUG \\
5003.55737 & -26.23 &  16.86 & TUG \\
5077.41720 & -45.88 &  11.99 & TUG \\
5077.43293 & -40.41 &  12.91 & TUG \\
5408.42261 & -48.03 &  11.29 & TUG \\
5408.44515 & -67.52 &  11.47 & TUG \\
5498.36262 &  22.78 &  18.37 & TUG \\
5498.38518 &   9.63 &  19.79 & TUG \\
5498.40774 &  14.99 &  19.26 & TUG \\
5841.43301 &  17.44 &  19.81 & TUG \\
5841.45487 &  35.24 &  18.46 & TUG \\
6088.45577 & -39.76 &  10.38 & TUG \\
6088.47848 &  -9.71 &  10.86 & TUG \\
6088.50104 & -33.56 &  10.75 & TUG \\
6181.39702 & -10.27 &  16.58 & TUG \\
6181.41970 & -27.56 &  14.90 & TUG \\
6473.55194 & -46.10 &  10.89 & TUG \\
6473.57450 & -34.12 &  13.38 & TUG \\
6483.44541 & -23.32 &  15.86 & TUG \\
6483.46797 & -39.55 &  13.53 & TUG \\
6591.25732 &   7.82 &  14.91 & TUG \\
6591.27985 &  14.77 &  15.38 & TUG \\
6826.49613 & -22.77 &  11.51 & TUG \\
6826.51867 & -15.76 &  10.73 & TUG \\
6877.43245 &  13.84 &  13.11 & TUG \\
6877.45501 & -13.28 &  13.95 & TUG \\
6967.25788 &  35.85 &  16.89 & TUG \\
6967.28043 &  23.79 &  17.20 & TUG \\
6967.30298 &  43.99 &  16.10 & TUG \\
7159.51955 & -39.39 &  18.36 & TUG \\
7159.54205 & -20.62 &  14.88 & TUG \\
7219.40794 & -23.15 &  21.11 & TUG \\
7219.43044 &   0.90 &  30.78 & TUG \\
7238.48844 &  38.55 &  21.30 & TUG \\
7265.27661 & -20.96 &  25.07 & TUG \\
7265.29911 &  -7.82 &  25.62 & TUG \\
7265.32161 & -24.20 &  22.20 & TUG \\
7265.40450 &  12.68 &  20.80 & TUG \\
7271.36935 & -22.67 &  14.77 & TUG \\
7271.39185 &   7.51 &  14.59 & TUG \\
7273.32616 &  18.43 &  10.27 & TUG \\
7273.34866 &  36.32 &  11.97 & TUG \\
7299.33026 &  60.47 &  20.94 & TUG \\
7299.35277 &  50.17 &  20.61 & TUG \\
7332.23936 &  38.92 &  11.52 & TUG \\
7332.26186 &   7.85 &  16.85 & TUG \\
7347.35361 &  67.23 &  14.84 & TUG \\
7351.22947 &  15.08 &  35.24 & TUG \\
7370.24002 &  79.20 &  27.69 & TUG \\
7370.26252 &  73.17 &  16.99 & TUG \\
7539.50055 & -17.34 &  14.61 & TUG \\
7539.52304 &   2.74 &  11.34 & TUG \\
7547.47297 &  27.53 &  17.72 & TUG \\
7547.49548 &   5.43 &  16.22 & TUG \\
7579.41323 & -29.49 &  16.99 & TUG \\
7581.39576 & -31.15 &  19.03 & TUG \\
7581.41828 &  13.32 &  18.11 & TUG \\
7647.37473 &  25.68 &  16.70 & TUG \\
7647.39723 &  40.75 &  27.66 & TUG \\
7651.44143 &  13.98 &  19.99 & TUG \\
7651.46394 &  -6.38 &  19.36 & TUG \\
7656.32711 &  37.59 &  17.23 & TUG \\
7656.34962 &  28.01 &  22.27 & TUG \\
7677.26921 &  40.41 &  14.30 & TUG \\
7677.29172 &  48.24 &  11.36 & TUG \\
7678.27693 &  52.54 &  13.71 & TUG \\
7678.29944 &  83.37 &  22.51 & TUG \\
7685.36031 &   9.16 &  21.95 & TUG \\
7685.38282 &  43.40 &  21.48 & TUG \\
7738.20224 &  35.41 &  17.77 & TUG \\
7738.22475 &  72.46 &  23.39 & TUG \\
7776.18465 &  40.48 &  19.06 & TUG \\
\hline
\end{tabular}\label{tab1}
\end{center}
\end{table}

\begin{table}{Table 1 Continued.} 
\scriptsize
\begin{center}
\begin{tabular}{cccc}
\hline
  BJD-2450000 [days]   & Velocity [$ms^{-1}$] & Uncertainty [$ms^{-1}$] & Remark \\
\hline
6887.23928 &  -7.67 &   3.90 & OAO \\
6915.08478 &   4.98 &   3.13 & OAO \\
6964.99969 &  27.05 &   3.72 & OAO \\
7003.95799 &  31.91 &   4.10 & OAO \\
7174.26960 &  -9.54 &   2.96 & OAO \\
7235.07821 & -12.65 &   3.36 & OAO \\
7250.08144 & -10.39 &   3.26 & OAO \\
7262.13763 &   1.81 &   3.20 & OAO \\
7284.09056 &  11.84 &   3.22 & OAO \\
7284.20209 &  14.86 &   3.20 & OAO \\
7307.04974 &  25.73 &   2.66 & OAO \\
7309.09047 &  25.54 &   3.74 & OAO \\
7331.04251 &  33.43 &   3.12 & OAO \\
7372.86249 &  33.99 &   3.90 & OAO \\
7403.87979 &  31.72 &   3.64 & OAO \\
7423.89012 &  41.73 &   9.39 & OAO \\
7475.33415 &   5.60 &   3.55 & OAO \\
7476.34130 &  -8.84 &   3.34 & OAO \\
7484.33302 &  -5.03 &   4.04 & OAO \\
7508.31346 &  -5.27 &   3.40 & OAO \\
7521.28512 &  -8.73 &   3.24 & OAO \\
7524.29154 & -14.12 &   6.53 & OAO \\
7542.29471 & -12.71 &   3.06 & OAO \\
7592.10450 &  -3.20 &   3.17 & OAO \\
7597.17800 &  -4.77 &   3.58 & OAO \\
7599.13811 & -11.39 &   3.00 & OAO \\
7625.21646 &  13.40 &   3.21 & OAO \\
7655.00139 &  31.92 &   3.92 & OAO \\
7666.97960 &  36.09 &   3.33 & OAO \\
7676.95567 &  31.52 &   4.01 & OAO \\
7688.95331 &  38.15 &   3.94 & OAO \\
7742.98101 &  49.39 &   3.72 & OAO \\
7756.92529 &  37.62 &   3.21 & OAO \\
7763.92901 &  32.44 &   3.55 & OAO \\
\hline
\end{tabular}
\end{center}
\end{table}

\section{Stellar properties}

\object{HD 208897} (\object{HIP 108513}) is a K0 giant star with a visual magnitude of $V=6.51$ and color index $B-V = 1.01$. The \textit{HIPPARCOS} \citep{vanleeuwen2007} parallax of 15.46 mas corresponds to a distance of 64.68 pc and the absolute visual magnitude obtained is $M_{V}=2.46$. The color excess $E(B-V)$ was estimated from the infrared dust emission maps of \citet{schlegel1998} and was calibrated according to \citet{beers2002}. Assuming the extinction to reddening ratio to be 3.1, the interstellar extinction was found to be at most $A_{V}=0.047$. The bolometric correction, $B.C = -0.392$, was taken from the \citet{flower96} tables. 

\begin{figure}
  \includegraphics[width=0.99\columnwidth]{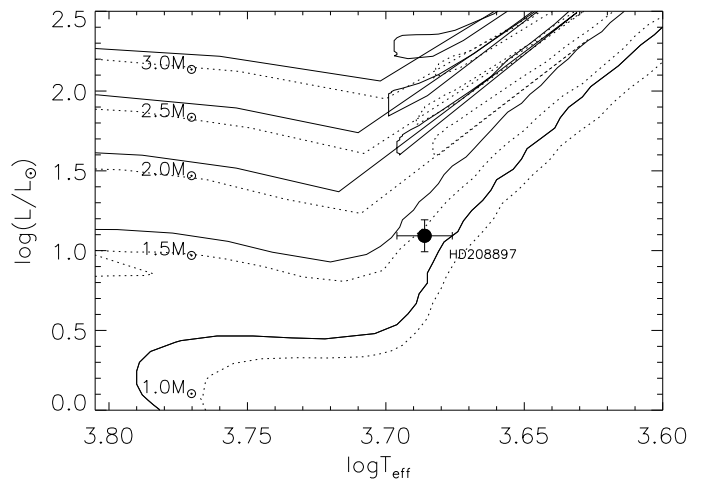}
  \caption{Location of \object{HD 208897} in the H-R diagram with evolutionary tracks \citep{lejeune2001} for Z = 0.008 (solid lines) and Z = 0.02 (dashed
lines) for masses of $1M_{\odot} - 3M_{\odot}$.}
\label{fig2}
\end{figure}

The stellar properties of \object{HD 208897} were derived using the equivalent width (EW) measurements of Fe I and Fe II lines from $I_{2}$-free spectrum taken with RTT150 CES. In the analysis, we excluded lines that were too weak ($<5$ m\AA) or too strong ($>100$ m\AA) lines and used the ODFNEW grid of Kurucz ATLAS9 model atmospheres \citep{castelli03}. In order to determine the stellar parameters, we iterated the stellar parameters ($T_{eff}$, $logg$, [Fe/H], microturbulance velocity $V_{t}$) with the help of the excitation/ionization balance of iron lines by calculating standard deviation of both $A(FeI)$\footnote{$A(FeI)=\log [N(FeI)/N(H)]+12$} and $A(FeII)$ abundances. With this method, we obtained the best solution with the iron abundance of $[Fe/H]=0.21 \pm0.15$ and microturbulent velocity of $v_{t}=1.28 \pm0.2$ kms$^{-1}$. This result indicates that \object{HD 208897} is a metal-rich star. The stellar atmosphere analysis yields the stellar parameters of $T_{eff}=4860 \pm100$ K, $vsini=3.9 \pm0.4$ kms$^{-1}$, and $logg=3.13 \pm0.1$. From the Stefan-Boltzmann law we derived the bolometric luminosity to be $L_{*}=12.3\pm1.1L_{\odot} $ and the stellar radius to be $R_{*}=4.98\pm0.2R_{\odot} $. We estimated the stellar mass to be $M_{*}=1.25M_{\odot} \pm0.1$ with the derived gravity and radius. The stellar parameters of the giant star \object{HD 208897} are summarized in Table \ref{tab2}, which lists those by \citet{wittenmyer2016} for comparison. The star positions in the Hertzsprung-Russell (H-R) diagram with the theoretical stellar isochrones are shown in Figure \ref{fig2}. It is clearly shown that the star has just begun to ascend the RGB. Based on the projected rotational velocity and star radius, we derived the upper limit for rotational period of 64 days for \object{HD 208897}.

\begin{table}
\caption{Stellar parameters of \object{HD 208897}.} 
\scriptsize
\begin{center}
\begin{tabular}{ccc}

\hline
  Parameter     & This work & \citet{wittenmyer2016} \\
\hline
  Sp. Type                 & K0                &          \\
  V $[mag]$                & 6.51              &          \\
  B-V                      & 1.01              &          \\
  $\pi$ $[mas]$            & $15.46 \pm 0.54$  &          \\
  B.C.                     & -0.392            &          \\
  $M_{V}$                  & 2.456             &          \\
  $A_{\nu}$                & 0.047             &          \\  
  $T_{eff}$ $[K]$          & $4860  \pm 100$   & 4905  \\
  log$L_{*}$ $[L_{\odot}]$  & $1.09  \pm 0.07$  & 1.09  \\
  log $g$ $[cgs]$          & $3.13  \pm 0.14$  & 3.38  \\
  $M_{*}$ $[M_{\odot}]$    & $1.25  \pm 0.11$  & 1.31  \\
  $R_{*}$ $[R_{\odot}]$    & $4.98  \pm 0.20$  & 4.88  \\
  $[Fe/H]$ $[dex]$         & $+0.21 \pm 0.15$  & +0.13 \\
  $v$sin$i$ $[kms^{-1}]$   & $3.90  \pm 0.42$  & -     \\
  $V_{t}$ $[kms^{-1}]$     & $1.28  \pm 0.24$  & 1.17  \\ 
\hline 
\end{tabular}\label{tab2}
\end{center}
\end{table}

\section{Results}

\subsection{Radial velocity and orbital solution}
\begin{figure}
  \includegraphics[width=0.99\columnwidth]{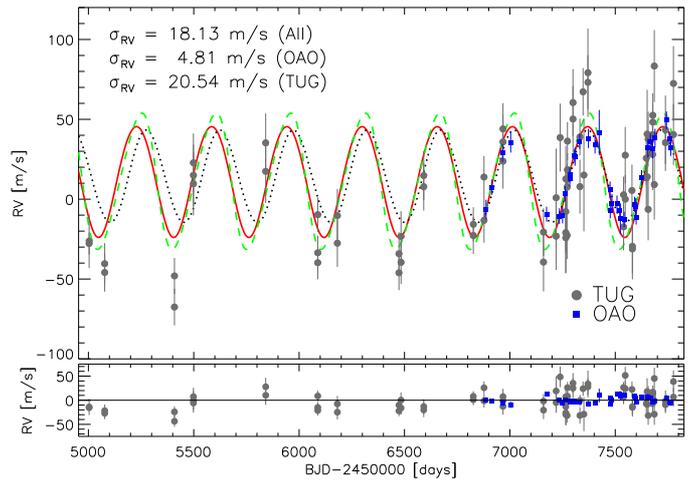}
  \caption{Observed RV variations and the best-fit orbital solutions of \object{HD 208897} with its residuals (bottom) to the best-fit-model. The solid red, dotted, and green dashed lines indicate the best Keplerian fit for TUG+OAO, OAO, and TUG, respectively. Dark gray points and blue squares represent data from TUG and OAO, respectively. The offset of 13.63 ms$^{-1}$ was introduced to HIDES data.  The RMS to the fits are 18.13 ms$^{-1}$, 4.81 ms$^{-1}$, and 20.54 ms$^{-1}$, respectively.}
  \label{fig3}
\end{figure}

We made the first observations of \object{HD 208897} at TUG and detected a significant RV variation. In order to verify RV variability, we started the OAO follow-up observations. We obtained 107 RV data in total, 34 of which were observed by HIDES at OAO. The observation dates, RVs, internal errors, and observation sites are listed in Table \ref{tab1}. Figure \ref{fig3} shows the observed RV curve. The dark gray circles and blue squares correspond to TUG and OAO, respectively. Both RV data show variability with an RMS scatter of 30-40 ms$^{-1}$, which is much larger than the measurements uncertainties and expected jitter levels for giant stars \citep{hekker2006}. Typical RV jitter for giant stars are at a level of around 10-20 ms$^{-1}$.

We performed the Lomb-Scargle (L-S) periodogram \citep{scargle1982} analysis for both TUG and OAO data, along with the whole data set, to search periodicity in the observed RV data. The L-S periodogram for the whole data set shows a significant periodicity at $\sim350$ days with a confidence level higher than 99.9\% (see Figure \ref{fig4} ) by calculating the false alarm probability (FAP). We estimated the FAP of the peak with the bootstrap randomization method. We created $10^5$ fake data sets by randomly redistributing the observed radial velocities while keeping the observation time fixed, and we subsequently applied the same periodogram analysis to these data sets. Only one fake data set showed a periodogram power higher than the observed data set. The periodic signal seen in the RV time series can be attributable to a planet that orbits around the star.

\begin{figure}
  \includegraphics[width=0.99\columnwidth]{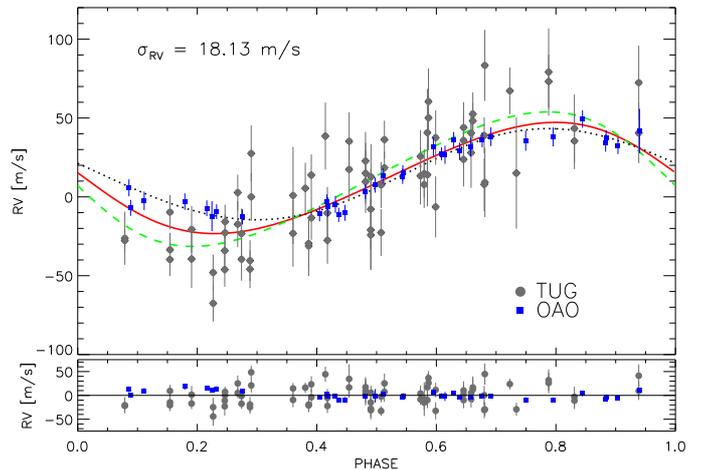}
  \caption{Same as Fig \ref{fig3} but shows observed RVs and best Keplerian fits at different orbital phases.}
  \label{fig3b}
\end{figure}

\begin{figure}
  \includegraphics[width=0.99\columnwidth]{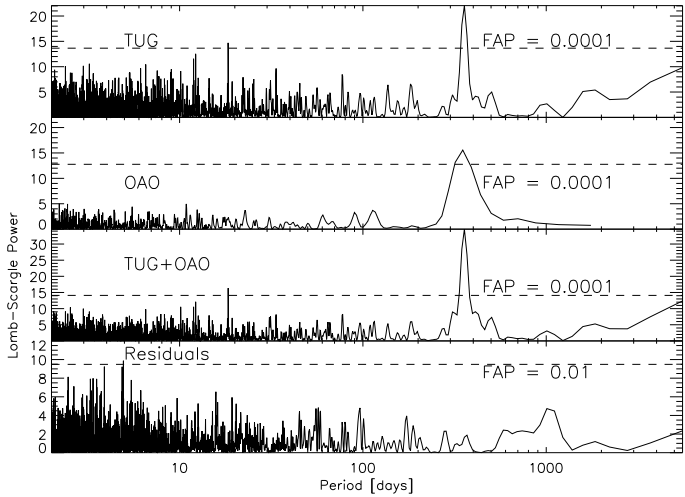}
  \caption{Lomb–Scargle periodograms for TUG, OAO, TUG+OAO RV measurements, and residuals to the Keplerian fit. The horizontal dotted lines indicate FAP thresholds. Possible peaks (FAP $\sim$ 0.0001) are seen at periods of about 350 days.}
  \label{fig4}
\end{figure}

We determined the parameter values of the Keplerian orbit by minimizing the \textbf{$\chi ^{2}$} statistic. We used the exofast \citep{eastman2013} IDL code,  which is based on MCMC algorithm.  We quadratically added stellar jitter ($\sigma^{2}_{error}=\sigma^{2}_{obs}+\sigma^{2}_{jitter}$) to intrinsic RV noise before performing the Keplerian fit to the RV data. We adopted optimal jitter value for the target when the reduced $\chi^{2}$ of the fit is close to unity. The adjustable parameters in the orbital solution are the orbital period $P$, time of periastron passage $T_{P}$, eccentricity $e$, velocity amplitude $K_{1}$, argument of periastron $w$, and RV zero-point $V_{0}$. The velocity offset was also applied between the TUG and OAO because we used two different stellar templates to derive the RV measurements for TUG and OAO observations and therefore this difference directly indicates a Doppler shift between the two templates. Figure \ref{fig3} and  \ref{fig3b} show the RV variations and the best Keplerian solutions for TUG+OAO (solid red line), OAO (dotted line), and TUG (solid green line) data, respectively. From the combined data, the radial velocities of \object{HD 208897} can be well fitted by an orbit with a period $P=352.7 \pm2$ days, a velocity semi-amplitude $K_{1}=34.7 \pm 2$ ms$^{-1}$, and an eccentricity $e=0.08 \pm0.06$. Adopting a stellar mass of 1.25$M_{\odot}$, we obtained a minimum mass for the companion as $m_{2}sini=1.40M_{J}$ and a semimajor axis as $a=1.05$ AU. The RMS of the residuals after subtraction of the best Keplerian fit is about 18.13 ms$^{-1}$ and this value is almost consistent with the measurement uncertainties for TUG data. The residuals to the Keplerian fit do not show a significant periodicity. When only OAO and TUG data were used, the minimum mass of the planet are derived to be $m_{2}sini=1.16M_{J}$ and $m_{2}sini=1.70M_{J}$, respectively. The RV precision of OAO data is higher than that of TUG, while the TUG observations cover a longer time span, hence allowing us to obtain more accurate periodicity for the orbit of companion (see Figure \ref{fig4}). The RV residuals from the best orbital fits of both data sets do not show any periodic variations. All uncertainties in the orbital analyses were derived using the bootstrap Monte Carlo approach by creating 1000 fake data sets. The best orbital parameters and physical properties of the proposed planet are summarized in Table \ref{tab3} with their errors.

\begin{table}
\caption{Orbital parameters of \object{HD 208897}.} 
\scriptsize
\begin{center}
\begin{tabular}{cccc}

\hline
Parameter       & TUG+OAO & OAO & TUG \\
\hline
P (days)                     \dotfill & 352.7 $\pm 1.7$ & 349.7 $\pm 3.3$ & 353.6 $\pm 2.7$ \\
$K_{1}$ (m$s^{-1}$)          \dotfill & 34.7  $\pm 2.2$ & 28.9  $\pm 1.2$ & 42.7  $\pm 5.5$  \\
e                            \dotfill & 0.07   $\pm 0.06$ & 0.04   $\pm 0.03$ & 0.15   $\pm 0.11$ \\
$\omega$ (deg)               \dotfill & 167     $\pm 83$   & 297    $\pm 64$   & 89     $\pm 42$  \\
$V_{0}$ (m$s^{-1}$)          \dotfill & 12.1   $\pm 1.8$  & 14.1   $\pm 0.9$  & 11.2   $\pm 3.8$  \\
$T_{p}$ (BJD-2450000)         \dotfill & 5036  $\pm 82$   & 6961 $\pm 54$ & 4971 $\pm 46$  \\
$m_{2}$ sin$i$ ($M_{J}$)     \dotfill & 1.40  $\pm 0.08$ & 1.16   $\pm 0.05$ & 1.70   $\pm 0.18$   \\
$a$ (AU)                     \dotfill & 1.05  $\pm 0.03$ & 1.04   $\pm 0.03$ & 1.05   $\pm 0.03$   \\
$f_{1}(m)$ $(10^{-9} M_{\odot})$   \dotfill & 1.5 $\pm 0.3$ & 0.8 $\pm 0.1$ & 1.7 $\pm 0.6$   \\
$a_{1}sini$ $(10^{-3} AU)$         \dotfill & 1.1 $\pm 0.1$ & 0.9 $\pm 0.2$ & 1.4 $\pm 0.3$   \\
$\sigma_{jitter}$ (m$s^{-1}$)      \dotfill & 12.0 & 4.0 & 12.0  \\
$\Delta RV$ (m$s^{-1}$)            \dotfill & 13.63 & - & -   \\
$N_{obs}$                          \dotfill & 107 & 34 & 73    \\
RMS (m$s^{-1}$)                    \dotfill & 18.13 & 4.81 & 20.54    \\
Reduced $\sqrt{\chi^{2}}$          \dotfill & 0.95 & 0.96 & 1.01    \\
\hline 
\end{tabular}\label{tab3}
\end{center}
\begin{center}
\noindent {\bf Note:} $\Delta RV$ offset between TUG and OAO velocities.
\end{center}
\end{table}

\begin{figure}
  \includegraphics[width=0.99\columnwidth]{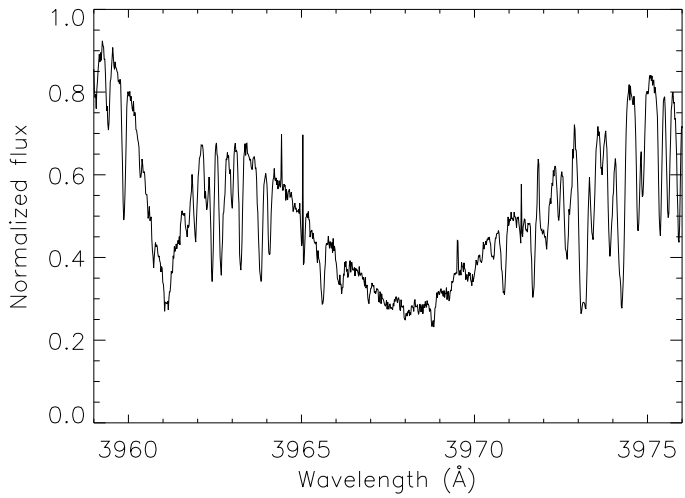}
  \caption{The Ca II H (3968.5{\AA}) absorption line region for \object{HD 208897}. The line core does not exhibit a significant emission.}
  \label{fig5}
\end{figure}

\begin{figure}
  \includegraphics[width=0.99\columnwidth]{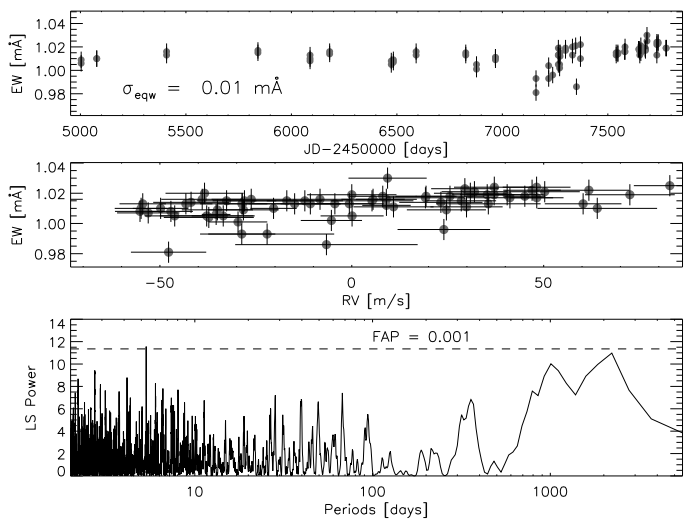}
  \caption{\textit{Top}: Variation of H$\alpha$ line EWs as a function of time. \textit{Middle}: EW measurements against the radial velocity. \textit{Bottom}: Periodogram of EW measurements. }
  \label{fig6}
\end{figure}

\begin{figure}
  \includegraphics[width=0.99\columnwidth]{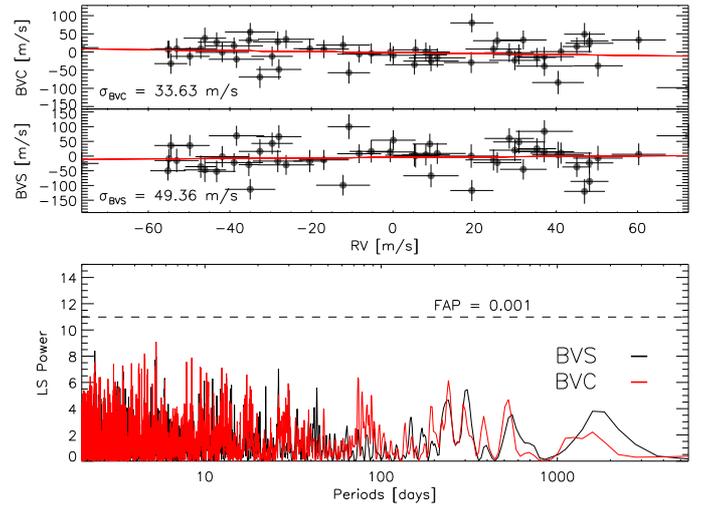}
  \caption{\textit{Top panel}: BVC and BVS variations for \object{HD 208897} The solid red line shows the linear fit of the bisectors. \textit{Bottom panel}: Periodograms of the bisectors of the CCFs; the red periodogram indicates BVC and black periodogram shows BVS. }
  \label{fig7}
\end{figure}

\begin{figure}
  \includegraphics[width=0.99\columnwidth]{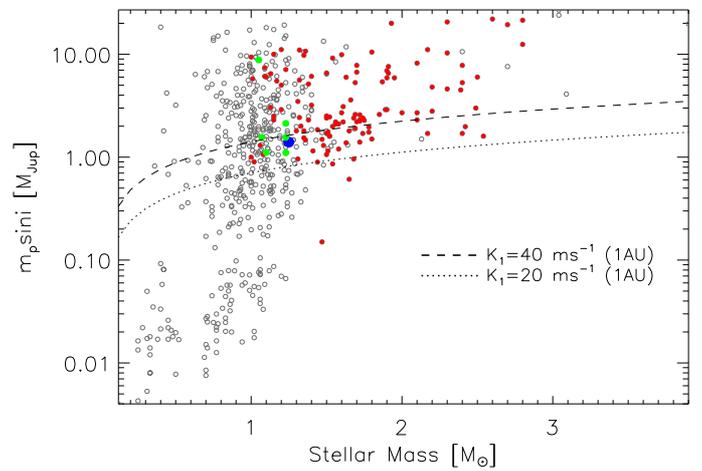}
  \caption{Planetary mass distribution against the mass of the host star. Filled red circles represent planetary systems orbiting intermediate-mass stars while blue filled circle indicates planet around \object{HD 208897}. Planets around less massive evolved intermediate-mass ($1.0M_{\odot}<M_{*}<1.3M_{\odot}$) and metal-rich ($[Fe/H]>0$) stars are plotted with green circles. Dashed and dotted lines correspond to the velocity semi-amplitude of 40 and 20 ms$^{-1}$ for a host star, respectively, imparted by a planet at 1 AU. }
  \label{fig8}
\end{figure}
 
\begin{figure}
  \includegraphics[width=0.99\columnwidth]{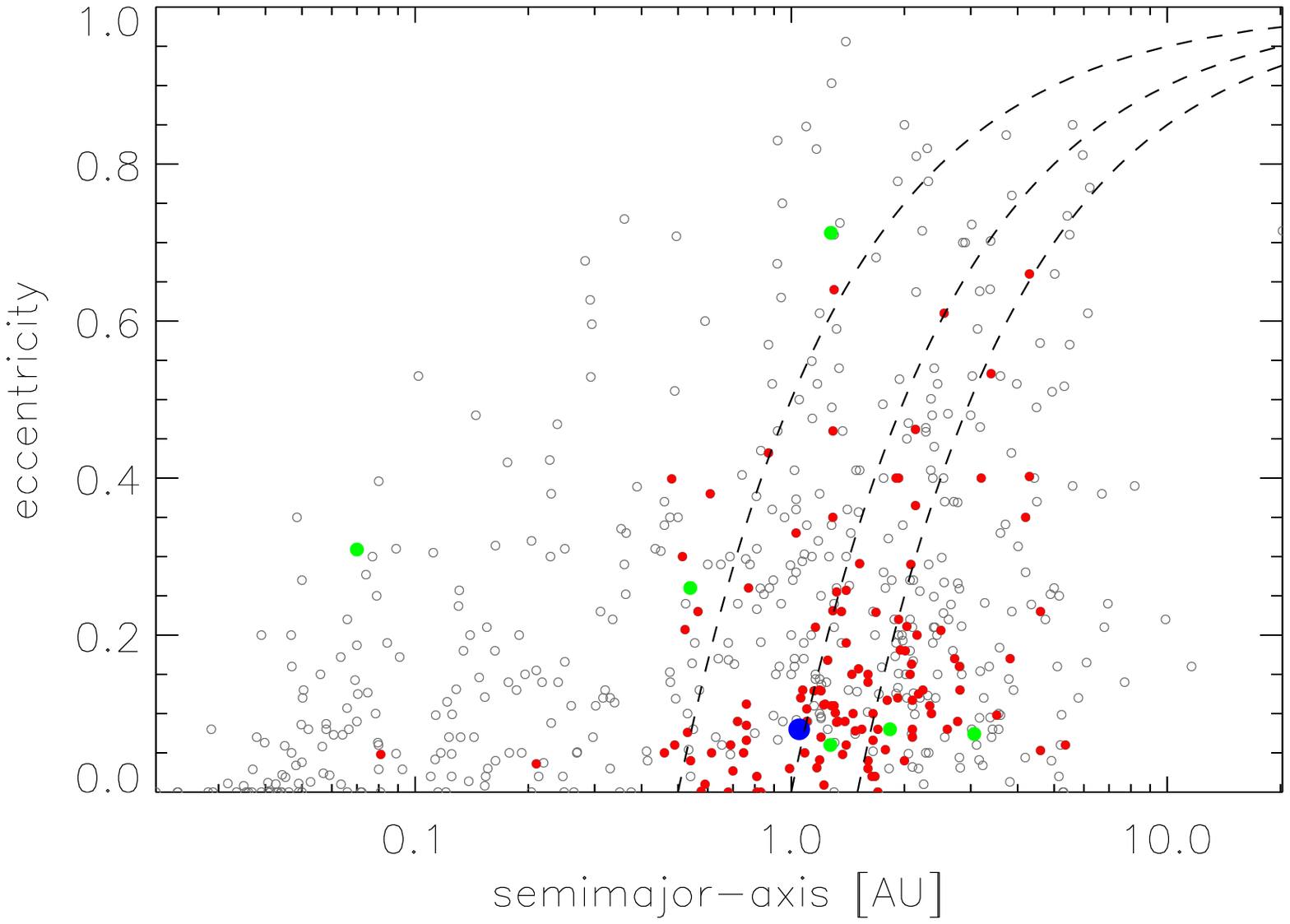}
  \caption{Planets distribution in the $a-e$ diagram. Planets around intermediate-mass stars are shown with filled red circles. The blue filled circle indicates the planetary system orbiting \object{HD 208897,} while green circles represent planets around $1.0M_{\odot}<M_{*}<1.3M_{\odot}$, {logg}$<4.0$ and metal-rich ($[Fe/H]>0$) stars. Dashed lines express the periastron distance [$q=a(1-e)$] of 0.5, 1.0, 1.5 AU, respectively, from the left. }
  \label{fig9}
\end{figure}

\subsection{Other mechanisms for tadial velocity variations}

There are many mechanisms, such as pulsation, inhomogeneous surface features, and stellar activities, that produce radial velocity variations and also create spectral line shape changes. To check these mechanisms, we examined the Ca II H and H$\alpha$ lines, photometric brightness variations, and spectral line shapes of \object{HD 208897}.

Activity induced hot or cool spots or plages on the surface of a cool star not only create asymmetries in the stellar absorption lines but also produce brightness variations caused by rotational modulations of these features. As can be seen in Figure \ref{fig1}, both AUKR photometric observations and \textit{HIPPARCOS} photometric measurements of \object{HD 208897} show that the star is stable and no significant correlation is seen in these data related to observed RV variation. According to RV amplitude-spot filling factor relation described by \citet{hatzes2002}, the observed RV amplitude of 35 ms$^{-1}$ would require a spot that covers 1.5\% of the stellar surface with a rotational velocity of about $vsini=4$ kms$^{-1}$. The expected photometric variability in this case is $\Delta m=0.05$ mag by assuming a temperature difference of $\Delta T=1200$ K between the spot and the stellar photosphere. This value  of variation is 1.5$\sigma$ above the observed scatter seen in the AUKR and \textit{HIPPARCOS} photometric data and can be easily detected. Also, from the projected rotational velocity and stellar radius we estimated an upper limit for the rotational period of $P_{rot}=64$ days, which is about five times smaller then the observed 352 days period of RV variation. Therefore we can quickly discard the hypothesis that rotational modulation creates RV variability in \object{HD 208897}.

The  Ca II H and K and H$\alpha$ line profiles are frequently used as the chromospheric activity indicators. As shown in Figure \ref{fig5}, the Ca II H line does not shows any emission feature at the line center. We only used HIDES spectra since TUG CES does not cover the spectral region where these lines are reside. We also measured EW of the H$\alpha$ line via a band pass of 1.0 {\AA} centered on 6562.808 {\AA} with the help of TUG CES data. Figure \ref{fig6} shows variations of EW measurements of H$\alpha$ line as a function of time and against RV. The L-S periodograms are also shown in the bottom panel of Figure \ref{fig6}. Clearly, the plot shows no correlation between RV and the H$\alpha$ EW and no significant periodicity exist in the H$\alpha$ line. Also, \citet{isaacson2010} showed that \object{HD 208897} has a small activity-index value of $S_{HK}=0.124$, which is consistent with our results. These results reinforce the existence of a substellar companion. 

Stellar intrinsic activities, such as pulsation or rotational modulation, can cause a change in the absorption line shape. Such a change may be misinterpreted as a Doppler shift of the lines by an orbital motion of the star. In order to examine spectral line shape, we performed an analysis of line bisectors based on cross-correlation function (CCF), which gives an average spectral line of the observed star \citep{gray2005}. The bisector analyses were performed with the TUG data. In the analysis, we used spectral lines that were located outside the $I_{2}$ absorption region because the $I_{2}$ lines affected the stellar spectrum. The CCFs were created by applying a special mask that discarded all the blended lines in the stellar spectrum and we identified 27 spectral lines that were relatively deep ($>0.3$). The bisector line was obtained by combining bisector points ranging from the core toward the wings of the CCF profile. We defined three flux levels to calculate velocity spans of the bisector: $V_T$ top (between 65\% and 85\% of the line depth from top), $V_C$ central (35-55\% ), and $V_B$ bottom zones (5-25\% ). The bisector velocity span ($BVS$) measurements were performed using the velocity difference between $V_T$ and $V_B$ ($BVS=V_T-V_B$), and the bisector curvatures ($BVC$) were derived using the difference of the velocity spans of the upper half of the bisector and the lower half ($BVC=(V_T-V_C)-(V_C-V_B)$). No correlation was found either between the RV and the BVS variations or between the RV and the BVC variations, which means the RV variations are not associated with the bisector variations. Also, no significant periods were obtained from the L-S periodogram analysis; all trial periods have extremely small significance levels. Moreover, we calculated the fundamental periods of radial pulsation with the method of \citet{cox1972} and also solar-like oscillation with the relationships of \citet{kjeldsen1995}. We estimated that the radial pulsation period of \object{HD 208897} to be less than one day and solar-like oscillation period of 0.8 days with RV amplitude of 2.5 ms$^{-1}$. These periods are more than two orders of magnitude shorter than the observed period of the RV variations. All these results suggest that the observed RV variations are consistent with the planetary hypothesis. In Figure \ref{fig7}, we presented BVS and BVC curves and their L-S periodograms.

\section{Discussion and conclusions}
In this paper, we report the first planet discovery in the TUG precise Doppler survey. The variability of the RV data observed at both TUG and OAO revealed a periodic signal, which suggests the presence of an unseen and probably low-mass companion. The AUKR and the \textit{HIPPARCOS} photometric data sets of \object{HD 208897} indicate that main cause of the observed RV variation is not rotational modulation due to stellar surface inhomogeneities.  Based on the rotational velocity and the radius of the star, the expected upper limit for the rotational period is about 64 days, which is much smaller than the observed RV variation period of 352 days. Also, our bisector analysis showed that there are no correlations between BVS and RV or between BVC and RV. Moreover, from the examination of the activity indicator Ca II H and H$\alpha$ lines we could not find any evidence for chromospheric activity. The estimated period of the fundamental radial mode pulsations and of solar-like oscillations were both derived as a few days. These values are much smaller than the RV variation seen in our data. These results indicate that observed RV variation in \object{HD 208897} can be attributed to an unseen companion. With the estimated stellar mass of 1.25$M_{\odot}$ and combined RV data set (TUG+OAO), we have found a giant planet with the mass of 1.40$M_{J}$ at 1.05 AU around the evolved intermediate-mass star \object{HD 208897}.  We did not find any long-term RV trend in the residuals of our best-fit model. Our stellar parameters indicate that \object{HD 208897} is a metal-rich star that is at the base of RGB phase.

One of the most remarkable features of the \object{HD 208897} system is the mass of the host star. Intermediate-mass stars are those that have masses higher than that of the Sun (typically 1.3$M_{\odot}$-5.0$M_{\odot}$). However, the mass of \object{HD 208897} is slightly smaller than this typical value. Figure \ref{fig8} shows the planetary mass distribution as a function of stellar mass. All of the planets orbiting around the evolved stars (\textit{logg}$<4.0$) in this figure are shown with filled red circles. The planets around evolved stars with masses between 1.0$M_{\odot}$ and 1.3$M_{\odot}$, and $[Fe/H]>0$  are also shown with green circles. Our results indicate that \object{HD 208897} is slightly metal-rich and has a mass at the lower limit of intermediate-mass stars range that harbor a $\sim$1.5 Jupiter mass size planetary companion. It is generally more difficult to detect such less massive planets around giants because of the relatively larger stellar jitter. However, our discovery shows that we can detect such less massive planets even around GK giants if we perform long-term observations.

The planet around \object{HD 208897} has an almost circular orbit ($e\sim 0.1$) and semimajor axis of about 1 AU. Figure \ref{fig9} demonstrates the distribution of semimajor axes of planets versus orbital eccentricities. Filled red circles indicate planets orbiting evolved intermediate-mass stars, while green circles represent planets around metal-rich evolved stars with masses between 1.0$M_{\odot}$ and 1.3$M_{\odot}$. As can be seen from the figure, there are only a few planets orbiting metal-rich evolved stars in this mass range and one of these is \object{HD 208897}. Most of the planets discovered around the evolved metal-rich stars have an eccentricity below 0.2 and a semimajor axis $a>1$ AU. Our results on \object{HD 208897} seem to reinforce the statistics. These results may be expected for evolved giants since those stars have begun to ascend the RGB and tidal influences would not become important for such distance planets yet. 
Planets at small orbital separations in evolved giants may be engulfed by their host stars during the stellar evolution \citep{sato2008, villaver2009}. The stellar parameters of \object{HD 208897}, however, indicate that it is at the earliest phase of RGB evolution and still has a radius (only $\sim0.025AU$) that is too small to tidally influence a planet at $\sim$1AU. Therefore the lack of inner Jupiter size planets or more massive planets than Jupiter in the star and the low eccentricity of the planet around \object{HD 208897} may be primordial.

Here, we have reported on one new planetary system around the evolved intermediate-mass star \object{HD 208897} based on precise RV measurements at TUG and OAO. This discovery will be important in understanding the planet formation around metal-rich intermediate-mass stars and the effect of stellar evolution on the planetary system configuration.

\begin{acknowledgements} This work was supported by The Scientific and Technological Research Council of Turkey (T\"{U}B\.{I}TAK), the project number of 114F099. Authors thank to TUG, KFU, and AST for partial support in using RTT150 (Russian-Turkish 1.5-m telescope in Antalya). This work was also supported by JSPS KAKENHI Grant Numbers JP23244038, JP16H02169. This work was funded by the subsidy 3.6714.2017 / 8.9 allocated to Kazan Federal University for the state assignment in the sphere of scientific activities.
\end{acknowledgements}

\bibliographystyle{aa} 
\bibliography{yilmaz_m_bibtex}

\end{document}